\documentclass[sigconf]{acmart}

\AtBeginDocument{%
  \providecommand\BibTeX{{%
    \normalfont B\kern-0.5em{\scshape i\kern-0.25em b}\kern-0.8em\TeX}}}

\copyrightyear{2024}
\acmYear{2024}
\acmDOI{XXXXXXX.XXXXXXX}

\acmConference[KDD'24]{}{August 25--29,
  2024}{Barcelona, Spain}
\acmISBN{978-1-4503-XXXX-X/18/06}

\usepackage{xcolor}
\usepackage{array}  
\usepackage{graphicx}
\usepackage{subcaption}
\usepackage[arrowdel]{physics}
\usepackage{algorithm}
\usepackage{algpseudocode}
\usepackage{bbm}
\usepackage{hyperref}
\hypersetup{
    colorlinks=true,
    linkcolor=blue,
    filecolor=magenta,      
    urlcolor=cyan,
}
\usepackage{float}
\usepackage{varwidth}
\usepackage{multirow}
\usepackage{subcaption}
\usepackage[export]{adjustbox}

\usepackage{todonotes}
\newcounter{todocounter}

\begin{document}

\title{Generalized User Representations for Transfer Learning} 

\author{Ghazal Fazelnia}
\email{ghazalf@spotify.com}
\orcid{orchid}
\affiliation{%
  \institution{Spotify}
  \city{New York}
  \state{NY}
  \country{USA}
}
\author{Sanket Gupta}
\email{sanketg@spotify.com}
\orcid{orchid}
\affiliation{%
  \institution{Spotify}
  \city{New York}
  \state{NY}
  \country{USA}
}
\author{Claire Keum}
\email{ckeum@spotify.com}
\orcid{orchid}
\affiliation{%
  \institution{Spotify}
  \city{New York}
  \state{NY}
  \country{USA}
}
\author{Mark Koh}
\email{markkoh@spotify.com}
\orcid{orchid}
\affiliation{%
  \institution{Spotify}
  \city{Golden}
  \state{CO}
  \country{USA}
}

\author{Ian Anderson}
\email{iananderson@spotify.com}
\orcid{orchid}
\affiliation{%
  \institution{Spotify}
  \city{New York}
  \state{NY}
  \country{USA}
}

\author{Mounia Lalmas}
\email{mounial@acm.org}
\orcid{orchid}
\affiliation{%
  \institution{Spotify}
  \city{London}
  \country{UK}
}

\begin{abstract}
We present a novel framework for user representation in large-scale recommender systems, aiming at effectively representing diverse user taste in a generalized manner. Our approach employs a two-stage methodology combining representation learning and transfer learning. The representation learning model uses an autoencoder that compresses various user features into a representation space. In the second stage, downstream task-specific models leverage user representations via transfer learning instead of curating user features individually. We further augment this methodology on the representation's input features to increase flexibility and enable reaction to user events, including new user experiences, in Near-Real Time. Additionally, we propose a novel solution to manage deployment of this framework in production models, allowing downstream models to work independently. We validate the performance of our framework through rigorous offline and online experiments within a large-scale system, showcasing its remarkable efficacy across multiple evaluation tasks. Finally, we show how the proposed framework can significantly reduce infrastructure costs compared to alternative approaches.  
\end{abstract}

\keywords{user model, recommender systems, cold-start models, autoencoders}


\maketitle
\pagestyle{plain}

\section{Introduction}
\label{Introduction}

Online music streaming services have become increasingly dominant in recent years. These services deliver catalogues with tens of millions of music tracks to hundreds of millions of users. While other online platforms utilize techniques that allow for personalized catalogue navigation and discovery of preferred content~\cite{zhang2019deep,amatriain2016past,bell2007lessons}, users frequently interact with music streaming services in a way that is distinct from other recommendation systems~\citep{schedl2018current,schedl2019deep,wang2018sequence,cebrian2010music,vall2017importance}.

Traditional user modeling captures and quantifies underlying user interest through explicit and implicit feedback~\citep{koren2009matrix,batmaz2019review,isinkaye2015recommendation,li2020survey}, but those systems are prone to scaling concerns when applied to the music domain. The robust catalogue is impacted by seasonality effects, exogenous events, and continuous additions of new music tracks that alter perceived relationships between tracks.
These tracks are often short and many tracks can be played together in a listening session, frequently without any feedback \citep{schedl2018current}. Users have conflicting interests to both revisit their favorite tracks and to discover new music to diversify their listening experiences \citep{zhang2012auralist, schedl2015tailoring}.
Furthermore, as interaction and consumption are the main sources of training for recommender systems, the lack of these signals make recommendations for new users an inherently challenging task~\citep{yurekli2021alleviating, wei2021contrastive}. Despite recent advancements in user modeling, capturing these user interests and modeling them remains a challenging task in large-scale music streaming services.

One approach is to treat each recommendation task independently: retrieval could be modeled directly on long-term taste; providing diverse listening experiences would be a multi-objective ranking problem; new users could use bandit style approaches from coarse early signals. However, these models would implicitly interact with one another without accounting for non-linear effects and each additional model raises the complexity of the infrastructure and overall cost. Furthermore, this can lead to disjoint experiences from the user's perspective as different models are engaged, which is especially important in those experiences early in a user's engagement with a streaming service \citep{passino2021next,DBLP:conf/recsys/Kula15}. 

As of writing, there is increased interest in so-called foundational models inspired by efforts in Natural Language Processing via large language models. These models are meant to serve myriad tasks with minimal adaptation to a novel domain. We take this concept and propose a new framework for user representation within an extremely large-scale industrial music recommendation system. In this framework, various downstream tasks continue to be developed independently while maintaining a consistent foundational user representation that is adaptable to new users and unseen tasks. By utilizing latent embeddings as a common interface, given their ease of efficient distribution, we are able to provide such a representation. We illustrate our results on a real-world music streaming dataset and demonstrate how each model component contributes to achieving the best result in modeling users. In this paper, we elaborate how our model builds an effective user representation, address the challenges implied by this framework, and demonstrate how this design can capture holistic user interests as well as current preferences.

{\textbf{Our contributions aim to answer the following research questions: }}
\begin{itemize}
    \item {\bf{[RQ1]}} How can we efficiently design a model to capture rich user representation in a large-scale recommender system that encompasses core user interests while being adaptable to various downstream tasks?
    \item {\bf{[RQ2]}} How can we design a user representation model that works for cold-start users and improves as users become more established on the platform? 
    \item {\bf{[RQ3]}} How can we devise effective evaluation strategies to measure efficacy of vector embedding space to ensure that we can provide value across a diverse range of downstream tasks?
\end{itemize}

We address these questions by designing a two-stage process. We develop an autoencoder model which takes in a variety of user features such as listening history, demographics, and contextual information. The main goal at this stage is to learn rich user representations that efficiently captures and summarizes user interests. To further adapt these representations for a variety of applications, we design the second stage to leverage learned representations in a transfer learning paradigm. Finally, we show the advantages of our method over the baselines in the empirical studies section.


\section{Related Works}

Modern recommender systems have been the core engine powering many user-facing platforms. Collaborative filtering methods have been one of the most successful approaches in the past few decades. By assuming that users with similar taste consume similar items, they are able to provide recommendations based on collaborative information. Similarly, matrix factorization methods have been successfully applied in many products and application across the industry~\citep{hu2008collaborative,mnih2008probabilistic}. Despite their efficiency, they are inherently linear, which limits their recommendation capabilities. Pivotal advances in deep learning have significantly enhanced the success of recommendation tasks over the years ~\citep{gopalan2015scalable,liang2018variational,xiao2015time,koren2009collaborative, zhou2019variational, sachdeva2019sequential, zhang2019deep}. 

Recommender systems provide personalized recommendations to users by learning from their implicit and/or explicit feedback~\citep{koren2009matrix,batmaz2019review,bell2007lessons,amatriain2016past}. Content-based algorithms leverage item-specific information while hybrid models combine collaborative and content-based approaches to further improve recommendations~\citep{aggarwal2016content, kouki2015hyper,ccano2017hybrid}. Autoencoder architectures and their variational adaptations are unsupervised methods that learn underlying themes in unlabeled data. In recommender systems this property is particularly desirable for understanding underlying user interests and learning fundamental user-item interaction patterns~\citep{liang2018variational,fazelnia2022variational}. While these models adeptly learn user interests in an unsupervised fashion, their inherent inflexibility for downstream tasks necessitates additional modifications to tailor them for specific recommendation objectives.


Transfer learning has become increasingly popular for recommender system applications in the last few years~\citep{fu2023exploring, yuan2021one}. This approach has shown promising results to address problems with insufficiency and sparsity in data through pre-training and fine-tuning stages~\citep{li2009transfer, pan2011transfer, pan2010transfer}. To learn robust user representations capable of capturing broad user interests and seamlessly integrating them into downstream applications, we have developed a transfer learning framework. This involves initially learning representations from user and track features and subsequently adapting them to a variety of downstream tasks. Unlike these recent works \citep{pancha2022pinnerformer, xia2023transact} that individually curate user features and require frequent retraining of the models, we take a fundamentally different approach by building a generalized user representation and then leveraging transfer learning for individual tasks. By making our embedding space stable, we reduce the retrain cadence to once every few months. 

These foundational models often tokenize inputs, leading to a vocabulary space with tens of thousands of items. However, music recommendation systems deal with extremely large-scale spaces consisting of tens of millions of tracks, which makes deploying these models extra challenging ~\citep{afchar2022explainability,schedl2018current,bonnin2014automated,cebrian2010music,vall2017importance}. To address this problem, we design a pre-processing step in which we learn embeddings for tracks through modality encoders. This step enables us to avoid using track IDs explicitly and significant improve the scalability of our framework.

The cold-start user problem is one the most challenging aspects of these user-facing services~\citep{yurekli2021alleviating,wei2021contrastive}. Lack of implicit or explicit signals from new users imposes new challenges to the traditional recommendation systems. Although cold-start recommendations is a well studied problem, most previous works focus on cold-start items \citep{panda2022approaches}. Methodologies have been proposed to incorporate external sources of information such as social networks connections \citep{nie2014information} or cross-domain fusions \citep{zhang2018crossrec}. While promising, these approaches cannot be applied in our real-world music streaming service due to user anonymity and privacy protection policies. In our framework, we incorporate platform onboarding signals for cold-start users alongside all features for established users into the model through near-real time updates and batch management. This enables us to significantly improve the results for cold-start users without compromising the performance for the established ones. 





\section{Approach}

In this section, we present the details of our approach by starting with introducing the notations. We describe the various components of our framework and summarize the training process for the user representation model. Finally, we present implementation details of the model.

\subsection{Notations}
Let $U$ be the set of users while $M$ is the set of tracks. 
For each user $u\in U$, we denote the features as $x_{u} \in\mathbf{R}^{d}$ where $d$ denotes the dimensionality of the input feature space. Let $z_u \in\mathbf{R}^{k}$ denote the user representation for user $u$ that exists in the representation space~$\Omega$. To learn and compress the information effectively, we set $k<<d$. 

Without loss of generality, we assume that we have a music streaming dataset consisting of user-track interactions. 
For each user $u$, $c_u$ is the concatenation of all user-specific information including context, demographics, affinities, and activities. More details regarding these features are presented in Section \ref{sec: implementation}.
$m_u$ is the set of all the tracks that user $u$ has interacted with. For each track $m$, we define track features as $t_m$. 

\subsection{Architecture}

\begin{figure}[t]
    \centering
    \includegraphics[width=0.45\textwidth]{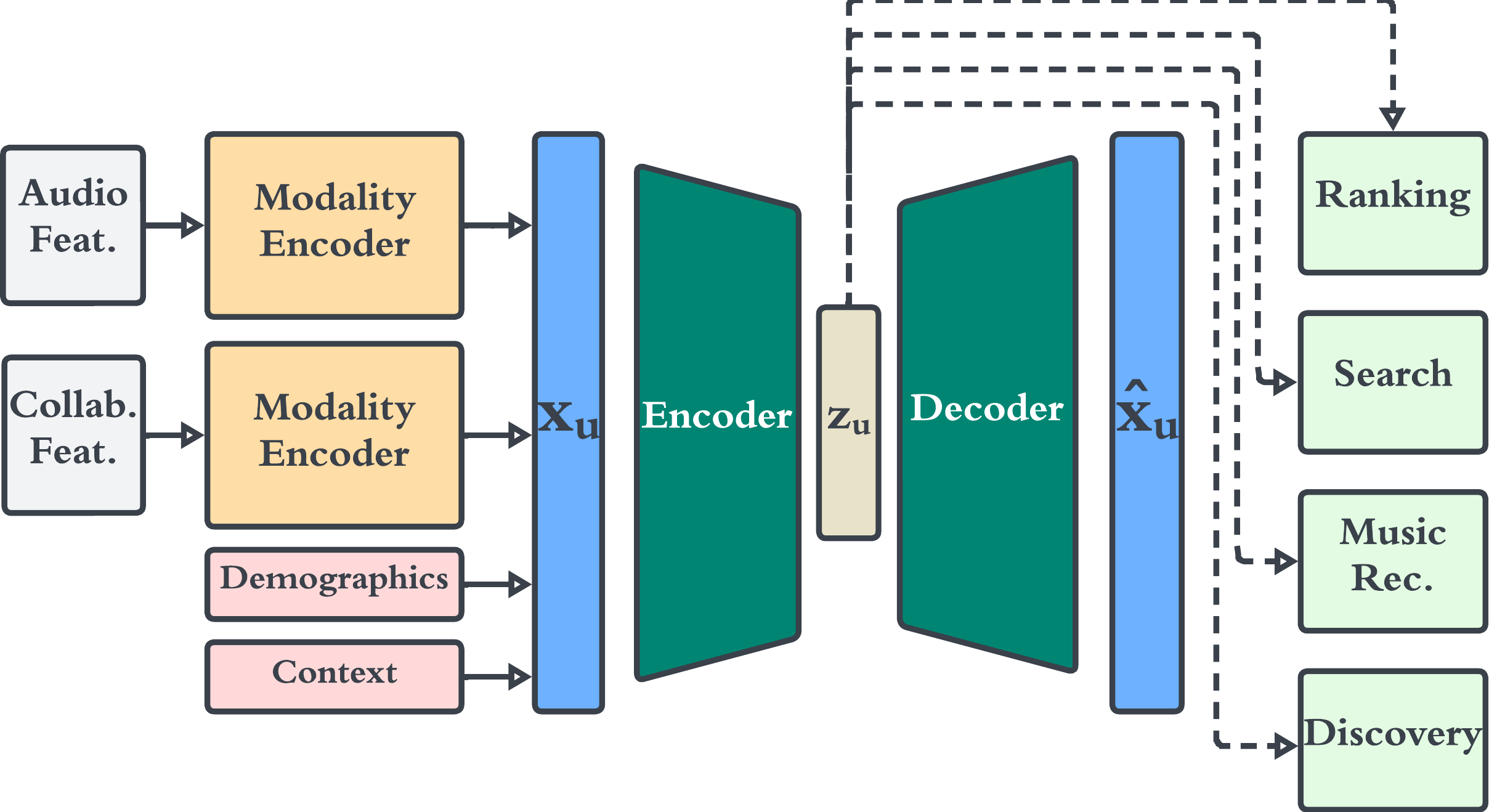}
    \caption{User Representation Model Architecture. Catalog interactions are embedded via pre-trained modality encoders to serve as input to an autoencoder model. The latent representation, $z_u$, serves as the User Representation for disparate downstream recommendation tasks.}
    \label{fig: model architecture}
\end{figure}

In this section, we present details of our user representation model architecture. Figure~\ref{fig: model architecture} presents the overview of the model architecture. Initially, modality encoders preprocess the track features as described below in Section~\ref{Modality_Encoders}. Subsequently, in conjunction with other user features, they form the input features $x$, and undergo the main model training to produce the user representations, $z$. These learned representations can then be leveraged as feature inputs suited to a diverse set of recommendation tasks. We provide more details for the last part in Section~\ref{sec: transfer learning methodology}.

\subsubsection{Modality Encoders}
\label{Modality_Encoders}
The first step of our modeling framework is to process the track space. As mentioned earlier, this step is necessary given the scale of the track space. Track audio is processed in a modality encoder component to produce an 80-dimensional real-valued vector mapping track into an audio embedding space. In addition, we assume if two tracks frequently appeared together in a playlist, they have latent similarities and are closer than two random tracks in an embedding space. To leverage this information, we let each track be represented by another 80-dimensional real-valued vector acquired in
a collaborative modality encoder. These track representations are based on track co-occurrences in playlists, meaning that two tracks are likely to be near each other in the embedding space if they co-occur in playlists and vice-versa. These particular embedding space has been previously shown to work well for music recommendation~\citep{mehrotra2018towards}. 


\subsubsection{Input Features}
After the pre-processing step with the modality encoders, we prepare the input features to the autoencoder model. We construct the input features for user $u$ and vectorize as follows:
\begin{equation}
    x_u=[c_u, a_u, v_u]
\end{equation}
$a_u$ and $v_u$ show the aggregate over audio embeddings and the collaborative embeddings of tracks consumed by user $u$, respectively.

\begin{algorithm}[t]
\caption{Training Generalized User Representation}\label{alg: training process}
\begin{algorithmic}
\State {\textbf{Input:}} Matrices $C$, $A$, $V$.
\While{not~~converged}
\State Sample a batch of users $\mathcal{D}$
    \State For $u \in \mathcal{D}$ construct $x_{u}$ as \\
    \State \;\;\;\; $x_u=[c_u, a_u, v_u]$ 
    \State \;\;\;\; $z_u = f_{enc}(x_u)$ \\
    \State Compute ${\mathcal{L}}_D = \frac{1}{D}\sum_{u=1}^{D} \| x_u-f_{dec}(f_{enc}(x_u))\|$

    \State Compute the aggregate gradient from this batch
    \State Update $f_{enc}$ and $f_{dec}$ by taking the gradient step.

\EndWhile
\State {\textbf{Output:}} $f^*_{enc}$ and $f^*_{dec}$
\end{algorithmic}
\end{algorithm}

\subsubsection{Autoencoder for Core User Representation}
Our model consists of an encoder and a decoder that together compress the user information into a representation space. The autoencoder architecture optimally learns from the input features and captures core user information into a representation space. The model aims to find $z_u \in\mathbf{R}^{k} $ such that it minimizes the loss function:
\begin{equation}\label{eq: loss function}
    {\mathcal{L}} = \frac{1}{U}\sum_{u=1}^{U} \| x_u-f_{dec}(f_{enc}(x_u))\|
\end{equation}
where $z_u=f^*_{enc}(x_u)$ for the optimal encoder $f^*_{enc}$.

We summarize the overall training steps in Algorithm~\ref{alg: training process}. We use a stochastic optimization on random batch of users to train the autoencdoer model and produce user representations. We note $A=\{a_u,  \forall u \in U\}$ and $V=\{v_u,  \forall u \in U\}$. More details regarding the training will be presented in next section. 





\subsection{Model Implementation}\label{sec: implementation}
We present the implementation details of our model.  We also show how we make the user representation generalizable, which plays a key role in several downstream tasks. 

\subsubsection{Training Dataset}
The offline training dataset is created using internal streaming information from our global user base, and contains instances of 600M+ users on 30M+ items in the catalog. 

\subsubsection{Input features}\label{sec: input features} We want the input features to capture both short-term and long-term trends in music taste and allow models to learn on both simultaneously. Our features are designed so that the user representation is generalizable to capture holistic understanding of the user at any given time. With this in mind, our features are listed below: 
\begin{itemize}
    \item Modality encoder output embeddings over different time horizons. We aggregate these embeddings over 1 week, 1 month and 6 months. As per Section~\ref{Modality_Encoders}, we use two different kinds of modality encoder embeddings, an embedding that processes acoustic information of audio, which we supplements with embeddings that capture the information of playlist co-occurrence of tracks to learn collaborative interests. 
    \item We combine this with demographic information such as country of registration as well as information on devices used and activity such as number of track plays. 
    \item To account for cold-start signals of new users, we combine the above features with new user onboarding information of artists and language. 
\end{itemize}

\begin{figure}
    \centering
    \includegraphics[width=0.3\textwidth]{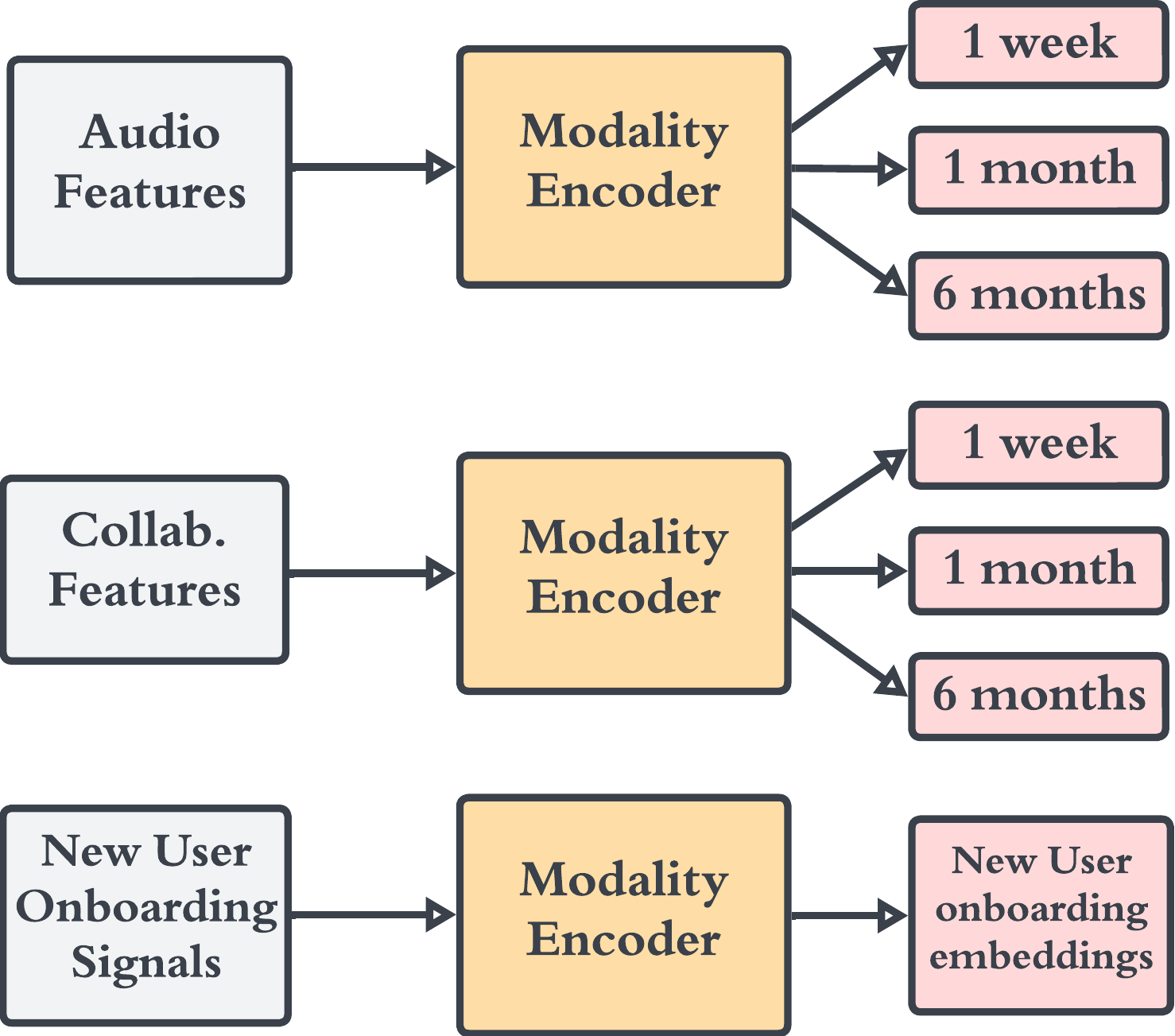}
    \caption{Embedding outputs of music modality encoders are aggregated over multiple time horizons.}
    \label{fig: embedding features}
\end{figure}

A key advantage of using multiple time frame based features is that our model can keep a “core” understanding of a user while being sensitive to recent taste changes. More research on this can be found in our previous work~\cite{fazelnia2022variational}. A diagrammed version of the embedding features are shown in Figure~\ref{fig: embedding features}. 

\begin{figure}[!htb]
    \centering
    \includegraphics[width=0.38\textwidth]{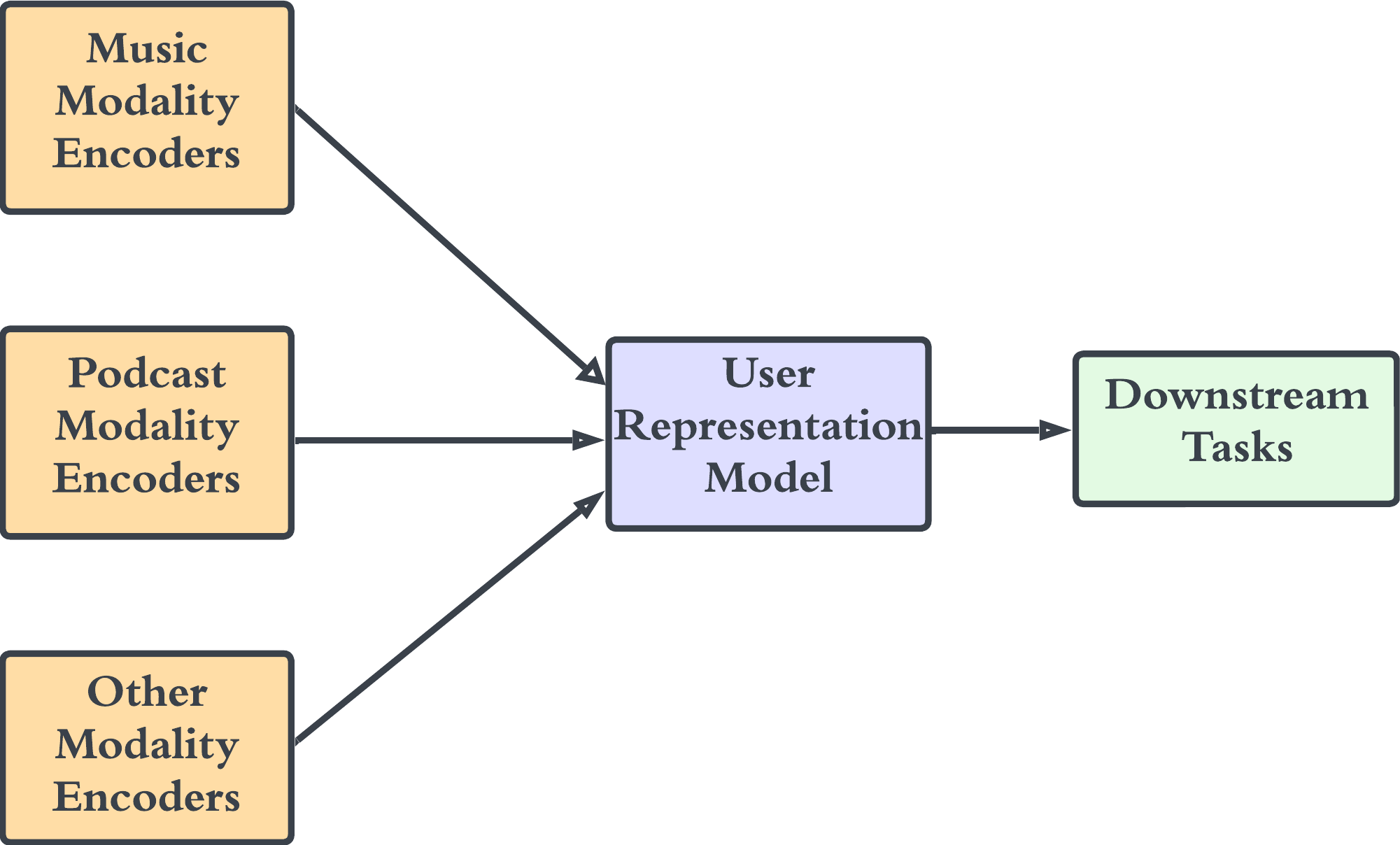}
    \caption{Music modality encoders as well as podcast modality encoders can feed into the user representation.}
    \label{fig: podcast features}
\end{figure}

This system allows us to accept not just the music streaming signals at different time frames but also supplement information from more modality encoders such as embeddings of other content information on the platform such as podcast listening. \footnote{Modern streaming platforms often host various content types including music and podcasts}
This allows the user representation to be universally useful across content types for recommendation tasks. 
Details of this can be seen in    Figure~\ref{fig: podcast features}.

\subsubsection{Model Details and Retrain Cadence}
The encoder and decoder of the autoencoder model are deep neural networks consisting of multiple hidden layers. We regularize by using a small dropout to ensure generalization and to avoid over-fitting. We use a scaled exponential linear unit (SELU) activation function in both encoder and decoder as it provides better convergence than other activation functions. 

Reconstruction loss allows us to update the weights of the model with the goal to summarize user features. This is in contrast with model architectures that prefer next item prediction. This ensures that we learn about the user holistically by summarizing them, instead of being just good at next action prediction. To ensure that we are holistically understanding the users, we evaluate user representation on multiple tasks including prediction of tracks listened in aggregate in a future time window. Details of this can be seen in Section 5.

The model outputs a 120-dimensional user representation that is available to be used for all downstream recommendation tasks. 

The model is trained in isolation from downstream tasks and is retrained once every few months. The schedule of its retraining is synchronized with upstream modality encoder retraining. Once the user model finishes retraining, downstream models perform their own retraining. More details on stability in the transfer learning model chain and batch synchronization is presented in Section~\ref{Batch_Synchronization_and_Management}. 

We combine both batch inference and near-real time inference for user representation. Batch inference pipelines run once daily for 600M+ users. Near-real time inference happens multiple times throughout the day depending on user activity.

\section{Transfer Learning Methodology} \label{sec: transfer learning methodology}
\begin{figure}[t]
    \centering
    \includegraphics[width=0.4\textwidth]{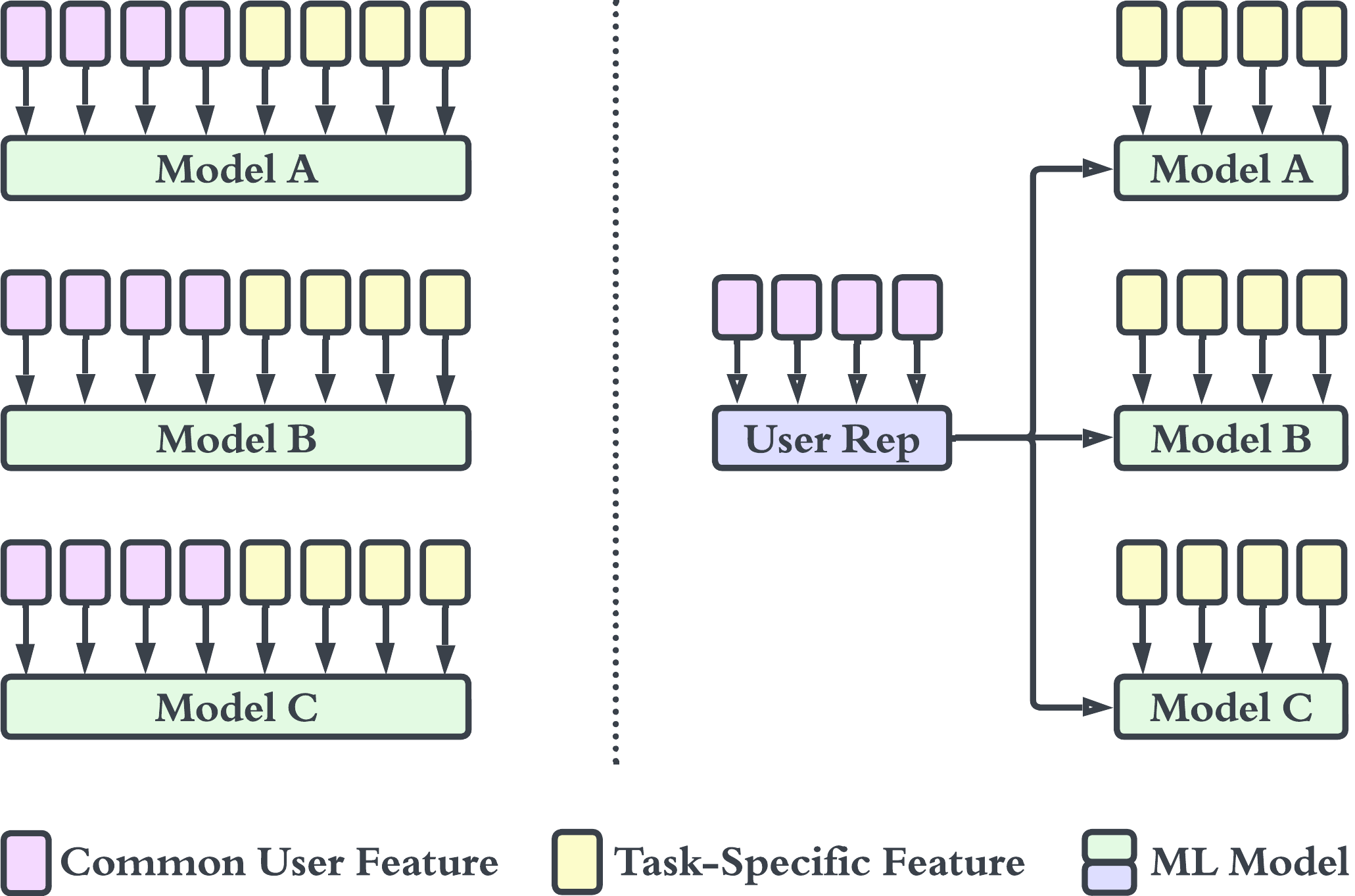}
    \caption{Downstream model tasks before (left) and after (right) incorporating transfer learning with a user representation.}
    \label{fig: transfer-learning-before-after}
\end{figure}

Transfer learning enables re-use of knowledge gained from a pre-trained model on new problems in downstream tasks such as ranking, search, music recommendation and discovery as shown in Figure~\ref{fig: model architecture}. This section presents our methodology of implementing transfer learning with user representation for large scale recommendation tasks. Figure \ref{fig: transfer-learning-before-after} depicts three tasks that require user information.  Without transfer learning (left) common user features are individually curated and fed into each model.  With transfer learning (right) common user features are instead condensed into a generalized user representation that can be fed directly into downstream task models.  By utilizing a generalized universal user representation instead of individual user features, we can reduce the amount of feature engineering and model complexity required in downstream models.

\subsection{Use Cases}
We present three production use cases of transfer learning in recommendation tasks, where user representation has provided value.

\paragraph{Classification models}

The first application is classification models. Classification models are really useful in tasks that require a system to know affinity or likelihood of engagement with a piece of content. An example of one such classifier is an artist preference model. It is a binary classifier that predicts a user's likelihood to follow an artist. The outputs of this model are used by teams that rank artists to show to users in cover art of playlists as well as in picking artists in playlist personalization tasks. User representation that captures user interest holistically and responds quickly to user interest is crucial to success. 

\paragraph{Candidate generation models}

The second application is candidate generation models. Candidate generation models are important in the first few stages of a recommender system and involve finding several items close to the user via nearest neighbor lookups. An example candidate generation model is a two-tower model that is used for picking items for recommendation \citep{covington2016deep}. The first tower takes user features and the second tower takes item features that are then passed through multiple hidden layers of a dense neural network. This model is tuned using dot product over the embeddings from the last layer of the two towers to bring them into the same vector space for nearest neighbor lookups in candidate generation tasks. User representation that is cold-start aware is important to make sure we are able to identify correct items to show to new users in their first few sessions.

\paragraph{Item ranking models}

The third application is item ranking models that are used to determine order of pieces of content in various parts of the application. An example is a ranking model that performs listwise ranking of items and re-orders them in a personalized order. Ranking models can have a simpler architecture while gaining all the capabilities of user representation. User representation that quickly responds to changing user taste is crucial in such models. \\

Given these use cases, there are specific qualities and characteristics of user representation which are of value. These are listed out in rest of the sub-sections.

\subsection{Responsiveness}

User representation should respond quickly to user listening behaviors and interactions. Batch inference systems take 2-3 days to respond to taste changes, which is not fast enough for a variety of downstream tasks. Representation at any point in time should reflect the latest information we have about user taste and continue to get updated with activity. To quickly respond to user listening behaviors and interactions, we use Near-Real Time (NRT) inference.

\begin{figure}[t]
    \centering
    \includegraphics[width=0.48\textwidth]{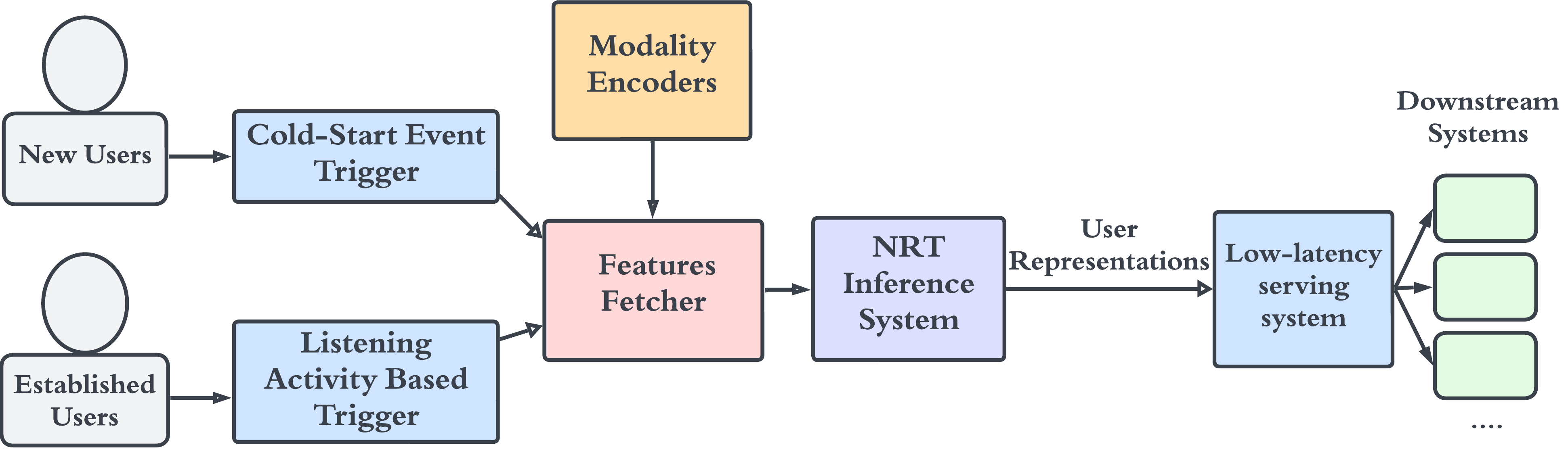}
    \caption{Near-Real Time Inference of User Representations. Cold-start user activations and user interactions trigger inferences, allowing for smooth updates to reflect the latest information.}
    \label{fig: nrt user model}
\end{figure}

Building out an event driven system to power NRT inference allows the representation to quickly respond to user taste. As shown in the Figure~\ref{fig: nrt user model}, we have event streams that act as trigger and carries input features based on users listening activities. This event queue is subscribed by our backend services that pre-process events into model consumable features and makes inferences in near-real time. The user representations are ingested into a low-latency serving system, which provides access to the most up-to-date representation to all downstream services. We combine both NRT and batch inference to make sure that we produce representation for active users as well as users who return to the platform after some inactivity. 


\subsection{Cold-Start Awareness}
\label{Impact_on_cold-start_Users}
User representation should work well for new users as much as established users, and should get better with time. On our platform, new users can select artists and/or language during the onboarding process. However, only a portion of new users complete the full onboarding process.

As shown in Figure~\ref{fig: nrt user model}, we use a separate event trigger with the onboarding signals. Artists selected during onboarding are passed to the modality encoder, which converts them to embeddings. Language selections are converted to a multi-hot encoded list. Embeddings for new users are created by the same model used for established users. The embeddings are available for immediate use by downstream models via our low-latency serving system. When new user onboarding signals are available, we use these alongside demographic features. In absence of new user onboarding signals, we impute these and use demographic and other static features. 

As users become more established on the platform, we keep inferring with onboarding selections alongside other listening history for a few months. After this point, new user onboarding signals no longer play a role in inference for a user. This is to ensure that personalization experience gracefully transitions from a user being cold-start to becoming established.



\subsection{Stability}
\label{Batch_Synchronization_and_Management}

In transfer learning, the outputs of an "upstream" model are fed into one or more "downstream" models, effectively forming a directed acyclic graph. For generality, note that these are general terms and a model can serve as both "downstream" to some models and "upstream" to others. In order for the user representation model to be generally usable for transfer learning, user representations and their respective modality encoders should exist in "stable" vector spaces.  A stable vector space is one in which item embeddings are only either added or updated after its initial training, without significantly changing the latent meaning of the individual vector space dimensions. In comparison to a vector space model which is retrained daily (in order to add and update items, for example), a stable vector space can be \textit{safely} interpreted by downstream models without needing to retrain the downstreams as new items come into existence. However in order to account for model drift, feature and hyperparameter updates, etc., even stable vector spaces require retraining~\citep{polyzotis2017skew}, albeit on a much less frequent cadence. Due to this, transfer learning introduces a challenge wherein changes to upstream models necessitate retraining downstream models to maintain compatibility. If an upstream model is retrained, downstream models must follow suit to avoid model failure or unpredictable outcomes due to data drift~\citep{polyzotis2017skew}. To address this, we propose "Batch Management," a strategy that ensures synchronization across models in a transfer learning chain.

\begin{figure}[t]
    \centering
    \includegraphics[width=0.6\linewidth]{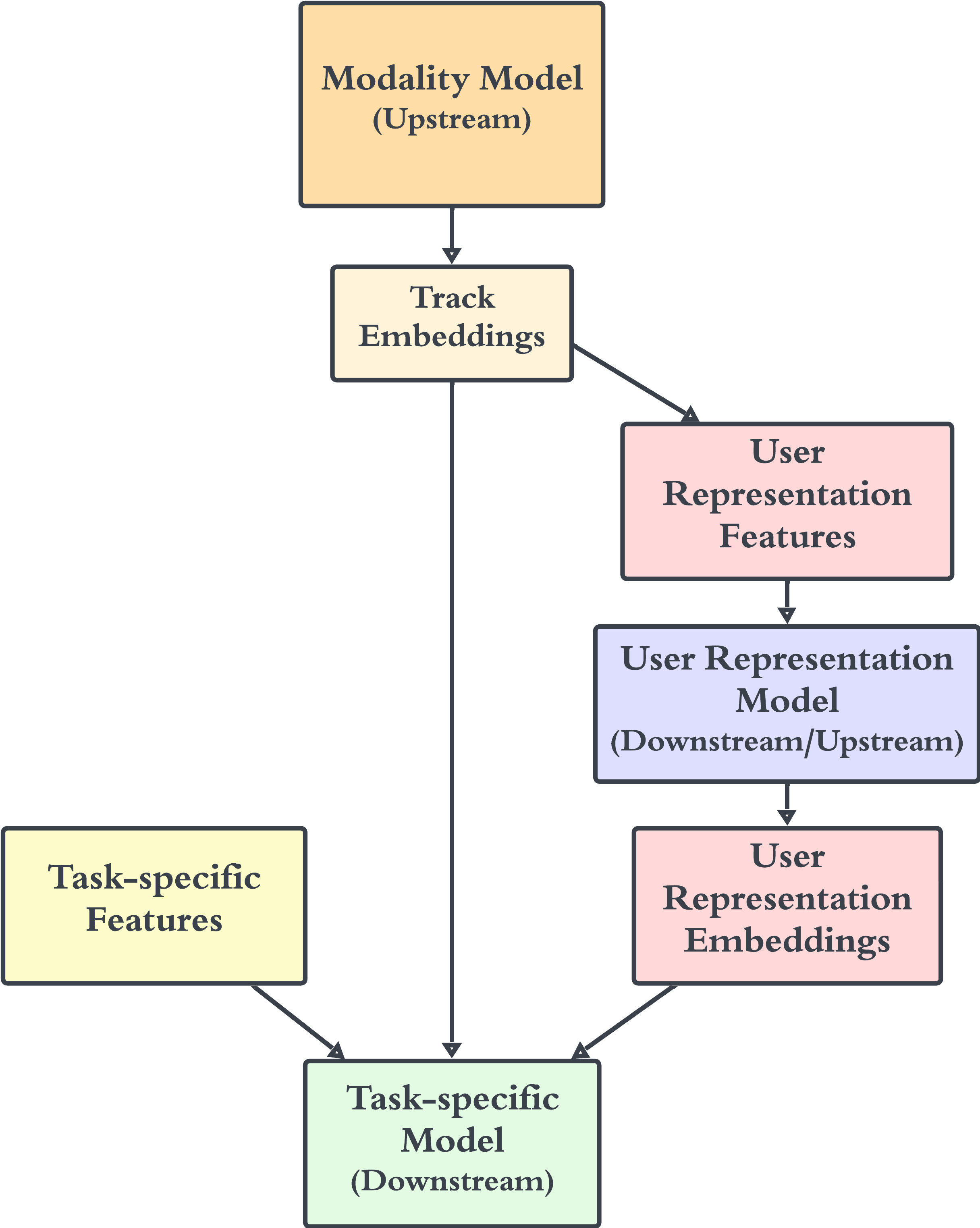}
    \caption{An example of a transfer learning model chain which requires "batch management".}
    \label{fig: transfer-learning}
\end{figure}

Figure~\ref{fig: transfer-learning} illustrates a common model chain: a modality encoder produces track embeddings, which are input into a user representation model. The resulting user representations are further fed into a task-specific downstream model alongside track embeddings and other task-specific features. Retraining the modality embedding model necessitates subsequent retraining of the user representation model and the task-specific model in that specific order. 

Key aspects of the "Batch Management" strategy are as follows:

    \begin{enumerate}
        \item[{1.}] \textbf{Batch Identification}: Retraining an embedding/vector space model results in a new "batch" with a tracked identifier allowing tracking of changes in vector space dimensions.
        \item[{2.}] \textbf{Synchronization}: Each model in the transfer learning chain is synchronized wrt. model retraining of its respective upstream models.
        \item[{3.}] \textbf{Upstream Retrains}: Models can be retrained independently as needed; however when any upstream model is retrained, all models which are downstream of it in the chain must automatically retrain.
        \item[{4.}] \textbf{Consistent Comparisons}: Downstream models should only compare embeddings within the same upstream batch(es).
        \item[{5.}] \textbf{Training and Inference}: Downstream models must use the same upstream batch(es) for both training and making predictions to ensure consistency.
        \item[{6.}] \textbf{Continuous Production}: Models in production can continue making up-to-date predictions during retraining.
    \end{enumerate}

To manage the complexity and resource consumption, we limit each model to two concurrent batches -- the "current" and "legacy" batches. "Batch rotation" occurs when a model completes training, pushing the new model in to the \textit{current} batch and rotating the previous batch to \textit{legacy}. After an upstream batch rotation, downstream models switch to using the legacy batch for offline and/or online inference while simultaneously retraining on the current batch. Once their retraining is complete, downstream models switch back to using the current batch for production.

This orchestration strategy accommodates various use cases, such as batch inference, online inference, and batch inference with online updates. While additional considerations exist for implementing Batch Management, they are beyond the scope of this~paper.

\section{Empirical Studies}
We present extensive evaluations to show the advantages of our framework on both cold-start users as well as established ones across multiple tasks and baselines. 
\subsection{Methodology}
To answer the aforementioned research questions in Section~\ref{Introduction}, we categorize our evaluation methodologies as follows:
\begin{enumerate}
 \item[{1.}] To answer {\bf{RQ1}} on capturing user interests while being adaptable to downstream tasks, we consider a fundamental recommendation task of future listening prediction. This task resembles the goals of many downstream models. We also evaluate on a downstream production model to verify our transfer learning hypothesis on a specific task.
 \item[{2.}] To answer {\bf{RQ2}} on capturing interests of cold-start users, we modify the future listening prediction task to reflect the timeline most important for cold-start users. We further breakdown the analysis for users who complete the onboarding process versus those who do not. 
 \item[{3.}] To answer {\bf{RQ3}} on devising effective evaluation strategies to measure efficacy of embedding space, we construct an intrinsic metric using cluster based evaluations. This ensures that user representation vector space is itself valuable to heterogeneous downstream models ranging from candidate generation to ranking.
\end{enumerate}


\subsubsection{Evaluations} We present three kinds of evaluations we used to verify our hypotheses. 
\paragraph{Future listening prediction}
The goal of this evaluation is to predict the tracks that a user listens to in the next $N$ hours or days. We set this evaluation to predict tracks in aggregate over $N$, instead of predicting next item or sequence of items. This is to ensure that we capture user interest holistically. We use user representation in a simple classifier model that is trained to predict future listening. 

We formulate this as a binary classification problem. We perform random negative sampling within the evaluation dataset to create equal number of positive and negative examples.

For established users, we investigate the performance in next 7 days to capture current interests holistically and in aggregate. For cold-start users, we investigate the performance in the next 4 hours after signing up. This is aimed at measuring the success of recommendations during the initial sessions following a user joining, as well as within a brief time frame while still in the cold-start mode. 

\paragraph{Production model downstream task}

To verify the advantage and efficacy of our model in a transfer learning case, we evaluate user representations on a real production model. We perform several offline experiments on this model and also compare infrastructure benefit of transfer learning.

\paragraph{Clustering based evaluations to find similar users}

\begin{figure}[t]
    \centering
    \includegraphics[width=0.35\textwidth]{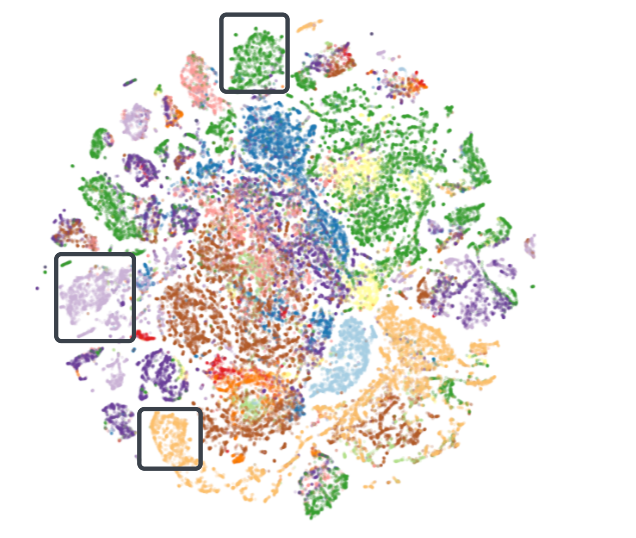}
    \caption{Intrinsic evaluation using nearest-neighbor lookups to find clusters of similar users in the User Representation space as visualized via a t-SNE plot.}
    \label{fig: clustering evals}
\end{figure}

This evaluation evaluates embeddings intrinsically using nearest neighbor lookup of users in representation space. We form clusters within this space and examine whether the representation can discern and capture them. We select random user samples and do approximate nearest neighbor retrieval to find top 50 neighbors. Then we measure how well these neighbors overlap with predefined clusters using normalized discounted cumulative gain (nDCG@50) as the metric. A simplified diagram of this evaluation can be seen in Figure~\ref{fig: clustering evals}.

%


\subsubsection{Data}
We create an offline training dataset using audio streaming information from the global user base. Our evaluation dataset consists of a few million users randomly sampled from the global user base. These users are sampled based on attributes such as country of registration, free vs premium status, as well as days since registration.

\subsection{Results}
We present the results of our empirical studies in this section. We break the section down by first presenting accuracy and AUC of ROC curves for established users and cold-start users on future listening prediction tasks. Next, we present offline and online results of downstream production model using user representation. Next, we present nDCG@50 of clustering based evaluations using various clustering heuristics. Finally, we present an ablation study of input features. 


\subsubsection{Future listening prediction}
We present the results of future track prediction for established users as well as cold-start users.  

\paragraph{Established users}
For established users, we use a 7 day window for future listening predictions. We compare user representation vs other popular and relatively scalable baselines for recommendation systems such as Non-Negative Matrix Factorization (NMF) \citep{zhang2006learning} and a deep learning based LightFM \citep{DBLP:conf/recsys/Kula15} model. 
We also compare performance of our user representation vs averaged embeddings of items in listening history using the same classifier.  As we can see in Table ~\ref{tab:results for established users}, we show improvements for established users in future 7d listening across various baselines. This helps us answer RQ1 with a fundamental task that resembles goals of downstream models. 

\begin{table}[t]
    \caption{Performance for predicting listening within next seven days for established users.}
    \centering
    \begin{tabular}{lcc}
        \toprule
        \textbf{Comparison} & \textbf{Accuracy} & \textbf{AUC} \\
        \midrule
        vs NMF & +15.2\% & +18.6\% \\
        vs LightFM & +26.2\% & +37.1\%  \\
        vs average embeddings & +1.8\% & +1.6\% \\
        \bottomrule
    \end{tabular}
    \label{tab:results for established users}
\end{table} 

\paragraph{Cold-start users}
Next, for cold-start users, we use a 4 hour window for future listening predictions. We first compare our user representation with a baseline of averaging the embeddings of artists selected during new user onboarding. Note that we do not have embeddings for cold-start users that do not complete onboarding process. For these users, we instead use the top popular tracks for the user's demographic. Table \ref{tab:results for cold-start users} shows the performance gains vs baseline for cold-start users. As can be seen, having a more unified approach helps with significant improvement in performance. This helps us answer RQ2 on whether we capture interests of cold-start users including those that complete and do not complete onboarding. 

\begin{table}[t]
    \caption{Performance for predicting listening within four hours following user registration.}
    \centering
    \begin{tabular}{>{\centering\arraybackslash}p{3cm}>
    {\centering\arraybackslash}p{2cm}>
    {\centering\arraybackslash}p{1cm}>{\centering\arraybackslash}p{1cm}}
        \toprule
        \textbf{Comparison} & \textbf{Onboarding Status} & \textbf{Accuracy} & \textbf{AUC} \\
        \midrule
        vs popularity heuristic & Completed & +26.2\% & +27.0\% \\
        vs popularity heuristic & Not Completed & +24.6\% & +24.7\% \\
        vs average embeddings &  Completed & +5.0\% & +5.1\% \\
        \bottomrule
    \end{tabular}

    \label{tab:results for cold-start users}
\end{table} 

\subsubsection{Downstream production model}
The downstream production model that we use for evaluation is an artist preference model: a classifier that predicts a user's likelihood to follow an artist. More details on this model are presented in Section 4. 

\paragraph{Offline Results}
For comparison, our baseline is a model that uses individually curated user specific features. These user features include several static, activity and listening history based features. We compare this with a model with the same training objective that instead uses leverages our user representation via transfer learning as shown in Figure~\ref{fig: transfer-learning-before-after}. As can be seen in Table~\ref{tab:table with downstream model results}, we show benefit of transfer learning with gains in AUC of 3.6\% when replacing all the individual user features with our user representation. \\

\begin{table}[t]
    \caption{Performance of a downstream model with user representation vs baseline.}
    \centering
    \begin{tabular}{lcc}
        \toprule
        \textbf{Model} & \textbf{Accuracy} & \textbf{AUC} \\
        \midrule
        Transfer Learning with User Rep & +3.8\% & +3.6\% \\
        \bottomrule
    \end{tabular}
    \label{tab:table with downstream model results}
\end{table} 
\vspace{-0.2cm}

\begin{figure}[t]
    \centering
    \includegraphics[width=0.42\textwidth]{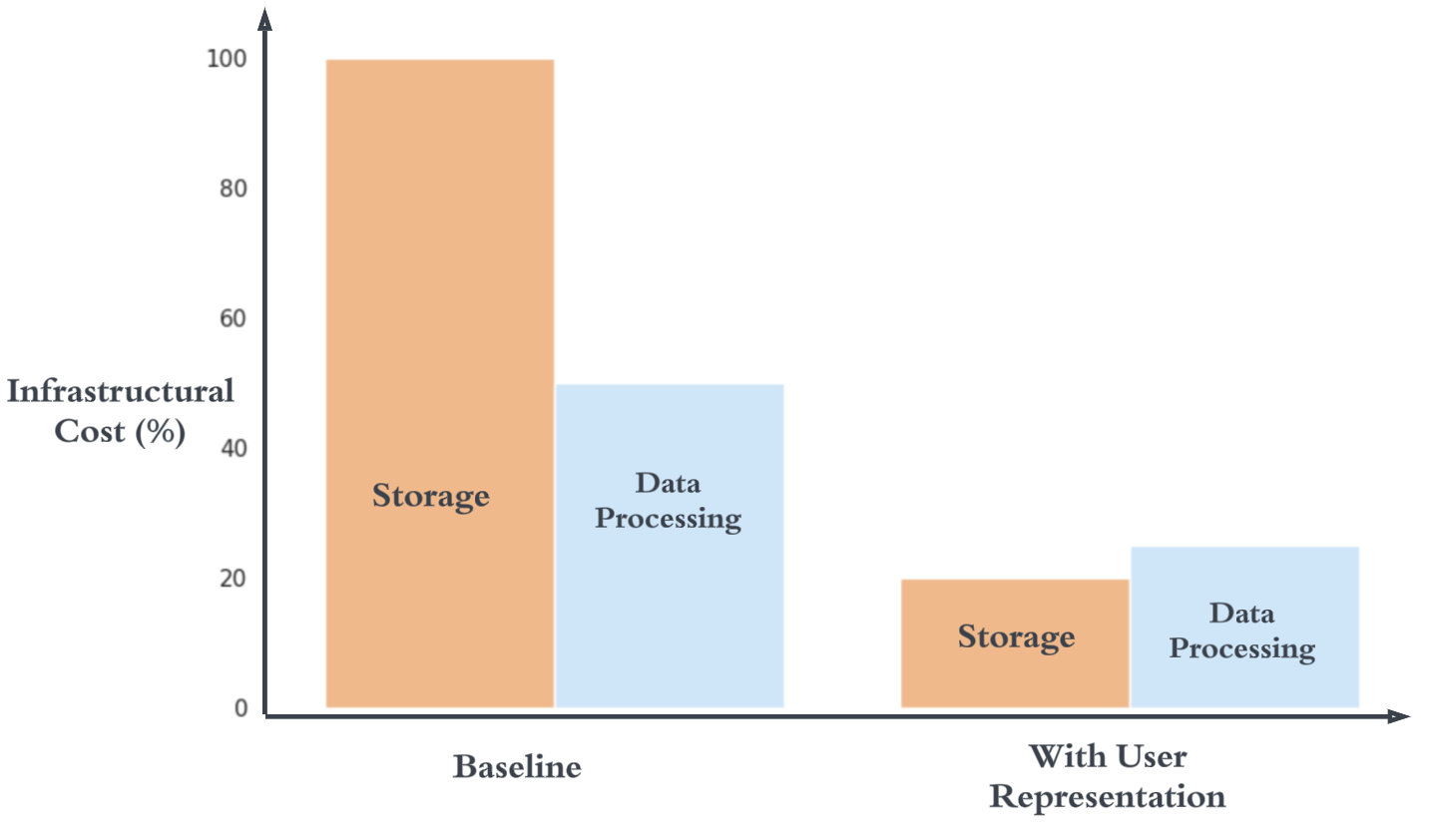}
    \caption{Relative infrastructural cost comparisons between system without and with user representation.}
    \label{fig: infrastructural benefit}
\end{figure}

\paragraph{Online Results}
We ran an online test with the same setup. For similar performance, we found significant infrastructure benefits including reduction in feature storage and data processing costs for inference as well as easier serving and on-call complexity. We were also able to leverage the existing serving systems and other centralized services that already exist in the ecosystem. This resulted in a large cost benefit, as shown in Figure~\ref{fig: infrastructural benefit}. These results help us further answer RQ1 by showing that we can capture user interests in a downstream production model. 

\subsubsection{Clustering based evaluations}
We build clusters of similar users based on known properties and use nearest neighbor lookups of subset of users to find overlap. We compare nDCG@50 for several types of clusters with average embeddings of items in users' listening histories. As we can see in Table ~\ref{tab:table with cluster evaluations}, we improve against the baseline across different kinds of user clusters. This helps us answer RQ3 by showing that representation space is intrinsically valuable.

\begin{table}[t]
    \caption{Performance at finding similar clusters for user representation vs. average embeddings.}
    \centering
    \begin{tabular}{lcc}
        \toprule
        \textbf{Cluster Heuristic} & \textbf{nDCG@50} \\
        \midrule
        Same favorite artists & +2.9\%  \\
        Same country of most listened artists & +5.5\% \\
        Same new user onboarding  & +26.2\% \\
        \bottomrule
    \end{tabular}
    \label{tab:table with cluster evaluations}
\end{table} 


\subsubsection{Ablation study}
Finally, to investigate relative feature importance we performed an ablation study on the input features. 
Baseline of the model is with features as laid out in Section~\ref{sec: input features}. We remove features to see the impact of them on the model.

\paragraph{Without new user onboarding signals}
Without new user onboarding cold-start signals, we see a decrease of 13.8\% in nDCG@50 with clusters made using same new user onboarding. Without onboarding signals, the model has a reduced ability to decipher new user clusters. 

\paragraph{Without modality encoder embeddings}
Without modality encoder embeddings, we see a decrease of 4.2\% in AUC of future listening evaluation over 7d. We also see a decrease of 37.1\% in nDCG@50 with clusters made with same favorite artists. These results show how modality encoder embeddings are crucial to overall success of the system.

\paragraph{Without user based static features}
Without user based static features including country of registration, we see a drop of 12.1\% in nDCG@50 with clusters made with same country of most listened artists.


This section presented our results to verify our hypotheses. To answer RQ1 and RQ2, we showed how a generalized user representation can help with future listening prediction tasks at various time windows for both established and cold-start users. We also presented offline and online results on a downstream production model, which showed large reductions in infrastructure costs. Finally, to answer RQ3, we showed the efficacy of the user representation in its embedding space using a variety of cluster evaluations. 
The generalized user representations have now been deployed in production and are being used to personalize content for our global user base. 

\section{Conclusion and Future Work}
In this paper, we presented a novel approach for learning generalized user representations in a large-scale music recommendation system. These representations capture core user interests and can be adapted to a variety of downstream tasks using transfer learning. To achieve the goal of being useful in downstream tasks, we make the user representation responsive in Near-Real Time (NRT) by adapting to changes in user taste quickly. We make the user representation cold-start aware by responding to onboarding signals alongside demographics in NRT and using the same model to infer both new and established users. Finally, we make the vector space stable to allow downstream models to have a retrain cadence of once every few months. We propose a novel batch management approach to enable stability of the vector space allowing downstream models to work independently. The effectiveness of our approach has been validated through a series of offline and online experiments.

While our focus has been on testing with music recommendation systems, there is potential to extend our approach to the broader audio domain, including podcast and audiobook recommendations. Beyond audio recommendations, other domains such as online news and e-commerce grapple with similar challenges in terms of scale, task heterogeneity, and addressing cold-start issues. Our framework presents an adaptable solution for such applications. Our emphasis has been primarily on track features using modality encoders. However, opportunities exist to incorporate additional information sources, including textual data such as lyrics, playlist and album titles. Given the recent advancements in Large Language Models (LLM), integrating embeddings generated by these models is another potential avenue to improve user representation.

\bibliographystyle{ACM-Reference-Format}

\bibliography{reference}


\begin{thebibliography}{46}


\ifx \showCODEN    \undefined \def \showCODEN     #1{\unskip}     \fi
\ifx \showDOI      \undefined \def \showDOI       #1{#1}\fi
\ifx \showISBNx    \undefined \def \showISBNx     #1{\unskip}     \fi
\ifx \showISBNxiii \undefined \def \showISBNxiii  #1{\unskip}     \fi
\ifx \showISSN     \undefined \def \showISSN      #1{\unskip}     \fi
\ifx \showLCCN     \undefined \def \showLCCN      #1{\unskip}     \fi
\ifx \shownote     \undefined \def \shownote      #1{#1}          \fi
\ifx \showarticletitle \undefined \def \showarticletitle #1{#1}   \fi
\ifx \showURL      \undefined \def \showURL       {\relax}        \fi
\providecommand\bibfield[2]{#2}
\providecommand\bibinfo[2]{#2}
\providecommand\natexlab[1]{#1}
\providecommand\showeprint[2][]{arXiv:#2}

\bibitem[Afchar et~al\mbox{.}(2022)]%
        {afchar2022explainability}
\bibfield{author}{\bibinfo{person}{Darius Afchar}, \bibinfo{person}{Alessandro Melchiorre}, \bibinfo{person}{Markus Schedl}, \bibinfo{person}{Romain Hennequin}, \bibinfo{person}{Elena Epure}, {and} \bibinfo{person}{Manuel Moussallam}.} \bibinfo{year}{2022}\natexlab{}.
\newblock \showarticletitle{Explainability in music recommender systems}.
\newblock \bibinfo{journal}{\emph{AI Magazine}} \bibinfo{volume}{43}, \bibinfo{number}{2} (\bibinfo{year}{2022}), \bibinfo{pages}{190--208}.
\newblock


\bibitem[Aggarwal(2016)]%
        {aggarwal2016content}
\bibfield{author}{\bibinfo{person}{Charu~C Aggarwal}.} \bibinfo{year}{2016}\natexlab{}.
\newblock \showarticletitle{Content-based recommender systems}.
\newblock In \bibinfo{booktitle}{\emph{Recommender systems}}. \bibinfo{publisher}{Springer}, \bibinfo{pages}{139--166}.
\newblock


\bibitem[Amatriain and Basilico(2016)]%
        {amatriain2016past}
\bibfield{author}{\bibinfo{person}{Xavier Amatriain} {and} \bibinfo{person}{Justin Basilico}.} \bibinfo{year}{2016}\natexlab{}.
\newblock \showarticletitle{Past, present, and future of recommender systems: An industry perspective}. In \bibinfo{booktitle}{\emph{Proceedings of the 10th ACM Conference on Recommender Systems}}. \bibinfo{pages}{211--214}.
\newblock


\bibitem[Batmaz et~al\mbox{.}(2019)]%
        {batmaz2019review}
\bibfield{author}{\bibinfo{person}{Zeynep Batmaz}, \bibinfo{person}{Ali Yurekli}, \bibinfo{person}{Alper Bilge}, {and} \bibinfo{person}{Cihan Kaleli}.} \bibinfo{year}{2019}\natexlab{}.
\newblock \showarticletitle{A review on deep learning for recommender systems: challenges and remedies}.
\newblock \bibinfo{journal}{\emph{Artificial Intelligence Review}} \bibinfo{volume}{52}, \bibinfo{number}{1} (\bibinfo{year}{2019}), \bibinfo{pages}{1--37}.
\newblock


\bibitem[Bell and Koren(2007)]%
        {bell2007lessons}
\bibfield{author}{\bibinfo{person}{Robert~M Bell} {and} \bibinfo{person}{Yehuda Koren}.} \bibinfo{year}{2007}\natexlab{}.
\newblock \showarticletitle{Lessons from the Netflix prize challenge}.
\newblock \bibinfo{journal}{\emph{Acm Sigkdd Explorations Newsletter}} \bibinfo{volume}{9}, \bibinfo{number}{2} (\bibinfo{year}{2007}), \bibinfo{pages}{75--79}.
\newblock


\bibitem[Bonnin and Jannach(2014)]%
        {bonnin2014automated}
\bibfield{author}{\bibinfo{person}{Geoffray Bonnin} {and} \bibinfo{person}{Dietmar Jannach}.} \bibinfo{year}{2014}\natexlab{}.
\newblock \showarticletitle{Automated generation of music playlists: Survey and experiments}.
\newblock \bibinfo{journal}{\emph{ACM Computing Surveys (CSUR)}} \bibinfo{volume}{47}, \bibinfo{number}{2} (\bibinfo{year}{2014}), \bibinfo{pages}{1--35}.
\newblock


\bibitem[{\c{C}}ano and Morisio(2017)]%
        {ccano2017hybrid}
\bibfield{author}{\bibinfo{person}{Erion {\c{C}}ano} {and} \bibinfo{person}{Maurizio Morisio}.} \bibinfo{year}{2017}\natexlab{}.
\newblock \showarticletitle{Hybrid recommender systems: A systematic literature review}.
\newblock \bibinfo{journal}{\emph{Intelligent Data Analysis}} \bibinfo{volume}{21}, \bibinfo{number}{6} (\bibinfo{year}{2017}), \bibinfo{pages}{1487--1524}.
\newblock


\bibitem[Cebri{\'a}n et~al\mbox{.}(2010)]%
        {cebrian2010music}
\bibfield{author}{\bibinfo{person}{Toni Cebri{\'a}n}, \bibinfo{person}{Marc Planagum{\`a}}, \bibinfo{person}{Paulo Villegas}, {and} \bibinfo{person}{Xavier Amatriain}.} \bibinfo{year}{2010}\natexlab{}.
\newblock \showarticletitle{Music recommendations with temporal context awareness}. In \bibinfo{booktitle}{\emph{Proceedings of the fourth ACM conference on Recommender systems}}. \bibinfo{pages}{349--352}.
\newblock


\bibitem[Covington et~al\mbox{.}(2016)]%
        {covington2016deep}
\bibfield{author}{\bibinfo{person}{Paul Covington}, \bibinfo{person}{Jay Adams}, {and} \bibinfo{person}{Emre Sargin}.} \bibinfo{year}{2016}\natexlab{}.
\newblock \showarticletitle{Deep neural networks for youtube recommendations}. In \bibinfo{booktitle}{\emph{Proceedings of the 10th ACM conference on recommender systems}}. \bibinfo{pages}{191--198}.
\newblock


\bibitem[Fazelnia et~al\mbox{.}(2022)]%
        {fazelnia2022variational}
\bibfield{author}{\bibinfo{person}{Ghazal Fazelnia}, \bibinfo{person}{Eric Simon}, \bibinfo{person}{Ian Anderson}, \bibinfo{person}{Benjamin Carterette}, {and} \bibinfo{person}{Mounia Lalmas}.} \bibinfo{year}{2022}\natexlab{}.
\newblock \showarticletitle{Variational user modeling with slow and fast features}. In \bibinfo{booktitle}{\emph{Proceedings of the Fifteenth ACM International Conference on Web Search and Data Mining}}. \bibinfo{pages}{271--279}.
\newblock


\bibitem[Fu et~al\mbox{.}(2023)]%
        {fu2023exploring}
\bibfield{author}{\bibinfo{person}{Junchen Fu}, \bibinfo{person}{Fajie Yuan}, \bibinfo{person}{Yu Song}, \bibinfo{person}{Zheng Yuan}, \bibinfo{person}{Mingyue Cheng}, \bibinfo{person}{Shenghui Cheng}, \bibinfo{person}{Jiaqi Zhang}, \bibinfo{person}{Jie Wang}, {and} \bibinfo{person}{Yunzhu Pan}.} \bibinfo{year}{2023}\natexlab{}.
\newblock \showarticletitle{Exploring Adapter-based Transfer Learning for Recommender Systems: Empirical Studies and Practical Insights}.
\newblock \bibinfo{journal}{\emph{arXiv preprint arXiv:2305.15036}} (\bibinfo{year}{2023}).
\newblock


\bibitem[Gopalan et~al\mbox{.}(2015)]%
        {gopalan2015scalable}
\bibfield{author}{\bibinfo{person}{Prem Gopalan}, \bibinfo{person}{Jake~M Hofman}, {and} \bibinfo{person}{David~M Blei}.} \bibinfo{year}{2015}\natexlab{}.
\newblock \showarticletitle{Scalable Recommendation with Hierarchical Poisson Factorization.}. In \bibinfo{booktitle}{\emph{UAI}}. \bibinfo{pages}{326--335}.
\newblock


\bibitem[Hu et~al\mbox{.}(2008)]%
        {hu2008collaborative}
\bibfield{author}{\bibinfo{person}{Yifan Hu}, \bibinfo{person}{Yehuda Koren}, {and} \bibinfo{person}{Chris Volinsky}.} \bibinfo{year}{2008}\natexlab{}.
\newblock \showarticletitle{Collaborative filtering for implicit feedback datasets}. In \bibinfo{booktitle}{\emph{2008 Eighth IEEE International Conference on Data Mining}}. Ieee, \bibinfo{pages}{263--272}.
\newblock


\bibitem[Isinkaye et~al\mbox{.}(2015)]%
        {isinkaye2015recommendation}
\bibfield{author}{\bibinfo{person}{Folasade~Olubusola Isinkaye}, \bibinfo{person}{YO Folajimi}, {and} \bibinfo{person}{Bolande~Adefowoke Ojokoh}.} \bibinfo{year}{2015}\natexlab{}.
\newblock \showarticletitle{Recommendation systems: Principles, methods and evaluation}.
\newblock \bibinfo{journal}{\emph{Egyptian informatics journal}} \bibinfo{volume}{16}, \bibinfo{number}{3} (\bibinfo{year}{2015}), \bibinfo{pages}{261--273}.
\newblock


\bibitem[Koren(2009)]%
        {koren2009collaborative}
\bibfield{author}{\bibinfo{person}{Yehuda Koren}.} \bibinfo{year}{2009}\natexlab{}.
\newblock \showarticletitle{Collaborative filtering with temporal dynamics}. In \bibinfo{booktitle}{\emph{Proceedings of the 15th ACM SIGKDD international conference on Knowledge discovery and data mining}}. \bibinfo{pages}{447--456}.
\newblock


\bibitem[Koren et~al\mbox{.}(2009)]%
        {koren2009matrix}
\bibfield{author}{\bibinfo{person}{Yehuda Koren}, \bibinfo{person}{Robert Bell}, {and} \bibinfo{person}{Chris Volinsky}.} \bibinfo{year}{2009}\natexlab{}.
\newblock \showarticletitle{Matrix factorization techniques for recommender systems}.
\newblock \bibinfo{journal}{\emph{Computer}} \bibinfo{volume}{42}, \bibinfo{number}{8} (\bibinfo{year}{2009}), \bibinfo{pages}{30--37}.
\newblock


\bibitem[Kouki et~al\mbox{.}(2015)]%
        {kouki2015hyper}
\bibfield{author}{\bibinfo{person}{Pigi Kouki}, \bibinfo{person}{Shobeir Fakhraei}, \bibinfo{person}{James Foulds}, \bibinfo{person}{Magdalini Eirinaki}, {and} \bibinfo{person}{Lise Getoor}.} \bibinfo{year}{2015}\natexlab{}.
\newblock \showarticletitle{Hyper: A flexible and extensible probabilistic framework for hybrid recommender systems}. In \bibinfo{booktitle}{\emph{Proceedings of the 9th ACM Conference on Recommender Systems}}. \bibinfo{pages}{99--106}.
\newblock


\bibitem[Kula(2015)]%
        {DBLP:conf/recsys/Kula15}
\bibfield{author}{\bibinfo{person}{Maciej Kula}.} \bibinfo{year}{2015}\natexlab{}.
\newblock \showarticletitle{Metadata Embeddings for User and Item Cold-start Recommendations}. In \bibinfo{booktitle}{\emph{Proceedings of the 2nd Workshop on New Trends on Content-Based Recommender Systems co-located with 9th {ACM} Conference on Recommender Systems (RecSys 2015), Vienna, Austria, September 16-20, 2015.}} \emph{(\bibinfo{series}{{CEUR} Workshop Proceedings}, Vol.~\bibinfo{volume}{1448})}, \bibfield{editor}{\bibinfo{person}{Toine Bogers} {and} \bibinfo{person}{Marijn Koolen}} (Eds.). \bibinfo{publisher}{CEUR-WS.org}, \bibinfo{pages}{14--21}.
\newblock
\urldef\tempurl%
\url{http://ceur-ws.org/Vol-1448/paper4.pdf}
\showURL{%
\tempurl}


\bibitem[Li et~al\mbox{.}(2009)]%
        {li2009transfer}
\bibfield{author}{\bibinfo{person}{Bin Li}, \bibinfo{person}{Qiang Yang}, {and} \bibinfo{person}{Xiangyang Xue}.} \bibinfo{year}{2009}\natexlab{}.
\newblock \showarticletitle{Transfer learning for collaborative filtering via a rating-matrix generative model}. In \bibinfo{booktitle}{\emph{Proceedings of the 26th annual international conference on machine learning}}. \bibinfo{pages}{617--624}.
\newblock


\bibitem[Li and Zhao(2020)]%
        {li2020survey}
\bibfield{author}{\bibinfo{person}{Sheng Li} {and} \bibinfo{person}{Handong Zhao}.} \bibinfo{year}{2020}\natexlab{}.
\newblock \showarticletitle{A Survey on Representation Learning for User Modeling.}. In \bibinfo{booktitle}{\emph{IJCAI}}. \bibinfo{pages}{4997--5003}.
\newblock


\bibitem[Liang et~al\mbox{.}(2018)]%
        {liang2018variational}
\bibfield{author}{\bibinfo{person}{Dawen Liang}, \bibinfo{person}{Rahul~G Krishnan}, \bibinfo{person}{Matthew~D Hoffman}, {and} \bibinfo{person}{Tony Jebara}.} \bibinfo{year}{2018}\natexlab{}.
\newblock \showarticletitle{Variational autoencoders for collaborative filtering}. In \bibinfo{booktitle}{\emph{Proceedings of the 2018 world wide web conference}}. \bibinfo{pages}{689--698}.
\newblock


\bibitem[Mehrotra et~al\mbox{.}(2018)]%
        {mehrotra2018towards}
\bibfield{author}{\bibinfo{person}{Rishabh Mehrotra}, \bibinfo{person}{James McInerney}, \bibinfo{person}{Hugues Bouchard}, \bibinfo{person}{Mounia Lalmas}, {and} \bibinfo{person}{Fernando Diaz}.} \bibinfo{year}{2018}\natexlab{}.
\newblock \showarticletitle{Towards a fair marketplace: Counterfactual evaluation of the trade-off between relevance, fairness \& satisfaction in recommendation systems}. In \bibinfo{booktitle}{\emph{Proceedings of the 27th acm international conference on information and knowledge management}}. \bibinfo{pages}{2243--2251}.
\newblock


\bibitem[Mnih and Salakhutdinov(2008)]%
        {mnih2008probabilistic}
\bibfield{author}{\bibinfo{person}{Andriy Mnih} {and} \bibinfo{person}{Russ~R Salakhutdinov}.} \bibinfo{year}{2008}\natexlab{}.
\newblock \showarticletitle{Probabilistic matrix factorization}. In \bibinfo{booktitle}{\emph{Advances in neural information processing systems}}. \bibinfo{pages}{1257--1264}.
\newblock


\bibitem[Nie et~al\mbox{.}(2014)]%
        {nie2014information}
\bibfield{author}{\bibinfo{person}{Da-Cheng Nie}, \bibinfo{person}{Zi-Ke Zhang}, \bibinfo{person}{Qiang Dong}, \bibinfo{person}{Chongjing Sun}, \bibinfo{person}{Yan Fu}, {et~al\mbox{.}}} \bibinfo{year}{2014}\natexlab{}.
\newblock \showarticletitle{Information filtering via biased random walk on coupled social network}.
\newblock \bibinfo{journal}{\emph{The Scientific World Journal}}  \bibinfo{volume}{2014} (\bibinfo{year}{2014}).
\newblock


\bibitem[Pan et~al\mbox{.}(2011)]%
        {pan2011transfer}
\bibfield{author}{\bibinfo{person}{Weike Pan}, \bibinfo{person}{Nathan~N Liu}, \bibinfo{person}{Evan~W Xiang}, {and} \bibinfo{person}{Qiang Yang}.} \bibinfo{year}{2011}\natexlab{}.
\newblock \showarticletitle{Transfer learning to predict missing ratings via heterogeneous user feedbacks}. In \bibinfo{booktitle}{\emph{Proceedings of the Twenty-Second International Joint Conference on Artificial Intelligence, Barcelona, Catalonia, Spain}}. \bibinfo{pages}{2318}.
\newblock


\bibitem[Pan et~al\mbox{.}(2010)]%
        {pan2010transfer}
\bibfield{author}{\bibinfo{person}{Weike Pan}, \bibinfo{person}{Evan Xiang}, \bibinfo{person}{Nathan Liu}, {and} \bibinfo{person}{Qiang Yang}.} \bibinfo{year}{2010}\natexlab{}.
\newblock \showarticletitle{Transfer learning in collaborative filtering for sparsity reduction}. In \bibinfo{booktitle}{\emph{Proceedings of the AAAI conference on artificial intelligence}}, Vol.~\bibinfo{volume}{24}. \bibinfo{pages}{230--235}.
\newblock


\bibitem[Pancha et~al\mbox{.}(2022)]%
        {pancha2022pinnerformer}
\bibfield{author}{\bibinfo{person}{Nikil Pancha}, \bibinfo{person}{Andrew Zhai}, \bibinfo{person}{Jure Leskovec}, {and} \bibinfo{person}{Charles Rosenberg}.} \bibinfo{year}{2022}\natexlab{}.
\newblock \showarticletitle{PinnerFormer: Sequence Modeling for User Representation at Pinterest}. In \bibinfo{booktitle}{\emph{Proceedings of the 28th ACM SIGKDD Conference on Knowledge Discovery and Data Mining}}. \bibinfo{pages}{3702--3712}.
\newblock


\bibitem[Panda and Ray(2022)]%
        {panda2022approaches}
\bibfield{author}{\bibinfo{person}{Deepak~Kumar Panda} {and} \bibinfo{person}{Sanjog Ray}.} \bibinfo{year}{2022}\natexlab{}.
\newblock \showarticletitle{Approaches and algorithms to mitigate cold start problems in recommender systems: a systematic literature review}.
\newblock \bibinfo{journal}{\emph{Journal of Intelligent Information Systems}} \bibinfo{volume}{59}, \bibinfo{number}{2} (\bibinfo{year}{2022}), \bibinfo{pages}{341--366}.
\newblock


\bibitem[Passino et~al\mbox{.}(2021)]%
        {passino2021next}
\bibfield{author}{\bibinfo{person}{Francesco~Sanna Passino}, \bibinfo{person}{Lucas Maystre}, \bibinfo{person}{Dmitrii Moor}, \bibinfo{person}{Ashton Anderson}, {and} \bibinfo{person}{Mounia Lalmas}.} \bibinfo{year}{2021}\natexlab{}.
\newblock \showarticletitle{Where To Next? A Dynamic Model of User Preferences.}
\newblock \bibinfo{journal}{\emph{WWW}}  \bibinfo{volume}{21} (\bibinfo{year}{2021}), \bibinfo{pages}{19--23}.
\newblock


\bibitem[Polyzotis et~al\mbox{.}(2017)]%
        {polyzotis2017skew}
\bibfield{author}{\bibinfo{person}{Neoklis Polyzotis}, \bibinfo{person}{Sudip Roy}, \bibinfo{person}{Steven~Euijong Whang}, {and} \bibinfo{person}{Martin Zinkevich}.} \bibinfo{year}{2017}\natexlab{}.
\newblock \showarticletitle{Data Management Challenges in Production Machine Learning}. In \bibinfo{booktitle}{\emph{Proceedings of the 2017 ACM International Conference on Management of Data}} (Chicago, Illinois, USA) \emph{(\bibinfo{series}{SIGMOD '17})}. \bibinfo{publisher}{Association for Computing Machinery}, \bibinfo{address}{New York, NY, USA}, \bibinfo{pages}{1723–1726}.
\newblock
\showISBNx{9781450341974}
\urldef\tempurl%
\url{https://doi.org/10.1145/3035918.3054782}
\showDOI{\tempurl}


\bibitem[Sachdeva et~al\mbox{.}(2019)]%
        {sachdeva2019sequential}
\bibfield{author}{\bibinfo{person}{Noveen Sachdeva}, \bibinfo{person}{Giuseppe Manco}, \bibinfo{person}{Ettore Ritacco}, {and} \bibinfo{person}{Vikram Pudi}.} \bibinfo{year}{2019}\natexlab{}.
\newblock \showarticletitle{Sequential variational autoencoders for collaborative filtering}. In \bibinfo{booktitle}{\emph{Proceedings of the Twelfth ACM International Conference on Web Search and Data Mining}}. \bibinfo{pages}{600--608}.
\newblock


\bibitem[Schedl(2019)]%
        {schedl2019deep}
\bibfield{author}{\bibinfo{person}{Markus Schedl}.} \bibinfo{year}{2019}\natexlab{}.
\newblock \showarticletitle{Deep learning in music recommendation systems}.
\newblock \bibinfo{journal}{\emph{Frontiers in Applied Mathematics and Statistics}}  \bibinfo{volume}{5} (\bibinfo{year}{2019}), \bibinfo{pages}{44}.
\newblock


\bibitem[Schedl and Hauger(2015)]%
        {schedl2015tailoring}
\bibfield{author}{\bibinfo{person}{Markus Schedl} {and} \bibinfo{person}{David Hauger}.} \bibinfo{year}{2015}\natexlab{}.
\newblock \showarticletitle{Tailoring music recommendations to users by considering diversity, mainstreaminess, and novelty}. In \bibinfo{booktitle}{\emph{Proceedings of the 38th international acm sigir conference on research and development in information retrieval}}. \bibinfo{pages}{947--950}.
\newblock


\bibitem[Schedl et~al\mbox{.}(2018)]%
        {schedl2018current}
\bibfield{author}{\bibinfo{person}{Markus Schedl}, \bibinfo{person}{Hamed Zamani}, \bibinfo{person}{Ching-Wei Chen}, \bibinfo{person}{Yashar Deldjoo}, {and} \bibinfo{person}{Mehdi Elahi}.} \bibinfo{year}{2018}\natexlab{}.
\newblock \showarticletitle{Current challenges and visions in music recommender systems research}.
\newblock \bibinfo{journal}{\emph{International Journal of Multimedia Information Retrieval}} \bibinfo{volume}{7}, \bibinfo{number}{2} (\bibinfo{year}{2018}), \bibinfo{pages}{95--116}.
\newblock


\bibitem[Vall et~al\mbox{.}(2017)]%
        {vall2017importance}
\bibfield{author}{\bibinfo{person}{Andreu Vall}, \bibinfo{person}{Massimo Quadrana}, \bibinfo{person}{Markus Schedl}, \bibinfo{person}{Gerhard Widmer}, {and} \bibinfo{person}{Paolo Cremonesi}.} \bibinfo{year}{2017}\natexlab{}.
\newblock \showarticletitle{The Importance of Song Context in Music Playlists.}. In \bibinfo{booktitle}{\emph{RecSys Posters}}.
\newblock


\bibitem[Wang et~al\mbox{.}(2018)]%
        {wang2018sequence}
\bibfield{author}{\bibinfo{person}{Dongjing Wang}, \bibinfo{person}{Shuiguang Deng}, {and} \bibinfo{person}{Guandong Xu}.} \bibinfo{year}{2018}\natexlab{}.
\newblock \showarticletitle{Sequence-based context-aware music recommendation}.
\newblock \bibinfo{journal}{\emph{Information Retrieval Journal}} \bibinfo{volume}{21}, \bibinfo{number}{2} (\bibinfo{year}{2018}), \bibinfo{pages}{230--252}.
\newblock


\bibitem[Wei et~al\mbox{.}(2021)]%
        {wei2021contrastive}
\bibfield{author}{\bibinfo{person}{Yinwei Wei}, \bibinfo{person}{Xiang Wang}, \bibinfo{person}{Qi Li}, \bibinfo{person}{Liqiang Nie}, \bibinfo{person}{Yan Li}, \bibinfo{person}{Xuanping Li}, {and} \bibinfo{person}{Tat-Seng Chua}.} \bibinfo{year}{2021}\natexlab{}.
\newblock \showarticletitle{Contrastive learning for cold-start recommendation}. In \bibinfo{booktitle}{\emph{Proceedings of the 29th ACM International Conference on Multimedia}}. \bibinfo{pages}{5382--5390}.
\newblock


\bibitem[Xia et~al\mbox{.}(2023)]%
        {xia2023transact}
\bibfield{author}{\bibinfo{person}{Xue Xia}, \bibinfo{person}{Pong Eksombatchai}, \bibinfo{person}{Nikil Pancha}, \bibinfo{person}{Dhruvil~Deven Badani}, \bibinfo{person}{Po-Wei Wang}, \bibinfo{person}{Neng Gu}, \bibinfo{person}{Saurabh~Vishwas Joshi}, \bibinfo{person}{Nazanin Farahpour}, \bibinfo{person}{Zhiyuan Zhang}, {and} \bibinfo{person}{Andrew Zhai}.} \bibinfo{year}{2023}\natexlab{}.
\newblock \showarticletitle{TransAct: Transformer-based Realtime User Action Model for Recommendation at Pinterest}.
\newblock \bibinfo{journal}{\emph{arXiv preprint arXiv:2306.00248}} (\bibinfo{year}{2023}).
\newblock


\bibitem[Xiao et~al\mbox{.}(2015)]%
        {xiao2015time}
\bibfield{author}{\bibinfo{person}{Yingyuan Xiao}, \bibinfo{person}{Pengqiang Ai}, \bibinfo{person}{Ching-Hsien Hsu}, \bibinfo{person}{Hongya Wang}, {and} \bibinfo{person}{Xu Jiao}.} \bibinfo{year}{2015}\natexlab{}.
\newblock \showarticletitle{Time-ordered collaborative filtering for news recommendation}.
\newblock \bibinfo{journal}{\emph{China Communications}} \bibinfo{volume}{12}, \bibinfo{number}{12} (\bibinfo{year}{2015}), \bibinfo{pages}{53--62}.
\newblock


\bibitem[Yuan et~al\mbox{.}(2021)]%
        {yuan2021one}
\bibfield{author}{\bibinfo{person}{Fajie Yuan}, \bibinfo{person}{Guoxiao Zhang}, \bibinfo{person}{Alexandros Karatzoglou}, \bibinfo{person}{Joemon Jose}, \bibinfo{person}{Beibei Kong}, {and} \bibinfo{person}{Yudong Li}.} \bibinfo{year}{2021}\natexlab{}.
\newblock \showarticletitle{One person, one model, one world: Learning continual user representation without forgetting}. In \bibinfo{booktitle}{\emph{Proceedings of the 44th International ACM SIGIR Conference on Research and Development in Information Retrieval}}. \bibinfo{pages}{696--705}.
\newblock


\bibitem[Y{\"u}rekli et~al\mbox{.}(2021)]%
        {yurekli2021alleviating}
\bibfield{author}{\bibinfo{person}{Ali Y{\"u}rekli}, \bibinfo{person}{Cihan Kaleli}, {and} \bibinfo{person}{Alper Bilge}.} \bibinfo{year}{2021}\natexlab{}.
\newblock \showarticletitle{Alleviating the cold-start playlist continuation in music recommendation using latent semantic indexing}.
\newblock \bibinfo{journal}{\emph{International Journal of Multimedia Information Retrieval}} \bibinfo{volume}{10}, \bibinfo{number}{3} (\bibinfo{year}{2021}), \bibinfo{pages}{185--198}.
\newblock


\bibitem[Zhang et~al\mbox{.}(2006)]%
        {zhang2006learning}
\bibfield{author}{\bibinfo{person}{Sheng Zhang}, \bibinfo{person}{Weihong Wang}, \bibinfo{person}{James Ford}, {and} \bibinfo{person}{Fillia Makedon}.} \bibinfo{year}{2006}\natexlab{}.
\newblock \showarticletitle{Learning from incomplete ratings using non-negative matrix factorization}. In \bibinfo{booktitle}{\emph{Proceedings of the 2006 SIAM international conference on data mining}}. SIAM, \bibinfo{pages}{549--553}.
\newblock


\bibitem[Zhang et~al\mbox{.}(2019)]%
        {zhang2019deep}
\bibfield{author}{\bibinfo{person}{Shuai Zhang}, \bibinfo{person}{Lina Yao}, \bibinfo{person}{Aixin Sun}, {and} \bibinfo{person}{Yi Tay}.} \bibinfo{year}{2019}\natexlab{}.
\newblock \showarticletitle{Deep learning based recommender system: A survey and new perspectives}.
\newblock \bibinfo{journal}{\emph{ACM Computing Surveys (CSUR)}} \bibinfo{volume}{52}, \bibinfo{number}{1} (\bibinfo{year}{2019}), \bibinfo{pages}{1--38}.
\newblock


\bibitem[Zhang et~al\mbox{.}(2018)]%
        {zhang2018crossrec}
\bibfield{author}{\bibinfo{person}{Yin Zhang}, \bibinfo{person}{Xiao Ma}, \bibinfo{person}{Shaohua Wan}, \bibinfo{person}{Haider Abbas}, {and} \bibinfo{person}{Mohsen Guizani}.} \bibinfo{year}{2018}\natexlab{}.
\newblock \showarticletitle{CrossRec: Cross-domain recommendations based on social big data and cognitive computing}.
\newblock \bibinfo{journal}{\emph{Mobile networks and applications}}  \bibinfo{volume}{23} (\bibinfo{year}{2018}), \bibinfo{pages}{1610--1623}.
\newblock


\bibitem[Zhang et~al\mbox{.}(2012)]%
        {zhang2012auralist}
\bibfield{author}{\bibinfo{person}{Yuan~Cao Zhang}, \bibinfo{person}{Diarmuid~{\'O} S{\'e}aghdha}, \bibinfo{person}{Daniele Quercia}, {and} \bibinfo{person}{Tamas Jambor}.} \bibinfo{year}{2012}\natexlab{}.
\newblock \showarticletitle{Auralist: introducing serendipity into music recommendation}. In \bibinfo{booktitle}{\emph{Proceedings of the fifth ACM international conference on Web search and data mining}}. \bibinfo{pages}{13--22}.
\newblock


\bibitem[Zhou et~al\mbox{.}(2019)]%
        {zhou2019variational}
\bibfield{author}{\bibinfo{person}{Fan Zhou}, \bibinfo{person}{Zijing Wen}, \bibinfo{person}{Kunpeng Zhang}, \bibinfo{person}{Goce Trajcevski}, {and} \bibinfo{person}{Ting Zhong}.} \bibinfo{year}{2019}\natexlab{}.
\newblock \showarticletitle{Variational session-based recommendation using normalizing flows}. In \bibinfo{booktitle}{\emph{The World Wide Web Conference}}. \bibinfo{pages}{3476--3475}.
\newblock


\end{thebibliography}
\newpage



\end{document}